\documentstyle[12pt]{article}
\begin{document}
\rightline{IMSc/94/38}
\rightline{\em Sept 94}
\baselineskip=24pt
\begin{center}
{\large Construction of  Yangian algebra  through a \\
 multi-deformation parameter dependent  rational
$R$-matrix}

\vspace {1.5 cm}
{\bf B. Basu-Mallick and P. Ramadevi}\\

The Institute of Mathematical Sciences, \\
C.I.T.Campus, Taramani, \\
Madras-600 113, INDIA. \\
{\it e-mail address: biru, rama@imsc.ernet.in }
\end{center}

\vspace {2.25 cm}
\noindent {\bf Abstract}

Yang-Baxterising a braid group representation associated with
 multideformed version of $GL_{q}(N)$ quantum group and
  taking  the corresponding $q\rightarrow 1$ limit, we obtain
  a rational $R$-matrix
which depends on $\left ( 1+ {N(N-1) \over 2} \right ) $
number of deformation parameters. By using  such rational $R$-matrix
subsequently we construct a multiparameter dependent extension
of  $Y(gl_N)$  Yangian algebra and find  that this extended  algebra
leads to   a modification  of  usual asymptotic  condition on  monodromy
matrix $T(\lambda )$, at   $ \lambda  \rightarrow \infty $ limit.
Moreover,  it turns out that,   there exists  a  nonlinear realisation of
this extended  algebra  through the generators of  original $Y(gl_N)$ algebra.
 Such   realisation interestingly  provides a
novel   $\left ( 1 + { N(N-1) \over 2 } \right ) $ number of
deformation parameter dependent coproduct  for
standard $Y(gl_N)$ algebra.

\newpage

\noindent \section{Introduction}
\setcounter{equation}{0}

 In recent years
 Yangian algebras [1] have  attracted much attention in the context of
 some $(1+1)$-dimensional integrable field models  having
nonlocal conserved quantities [2,3].
 Moreover,  it has been  observed
that a class of  one dimensional quantum  spin chains with long
ranged  interactions also possess this novel Yangian symmetry [4-7].
Such a  symmetry
enables one to derive  closed form expressions for many
physical quantities like thermodynamic potential and ground state
correlation function [7].

These  Yangian algebras  can   be constructed  by using
the well known quantum Yang-Baxter equation (QYBE) [1,2]
\begin{equation}
R(\lambda - \mu)~T_1(\lambda)~T_2(\mu)~=~
T_2(\mu)~T_1(\lambda)~R(\lambda-\mu) \, ,     \label{(1.1)}
\end{equation}
where $~ T_1(\lambda) =
T(\lambda) \otimes {\bf 1},~ T_2(\mu) = {\bf 1} \otimes T(\mu),$
and $T(\lambda )$
 is a  $N \times N $ operator valued matrix.  The
 $R(\lambda -\mu)$  appearing  in  QYBE
is taken as  a $N^2 \times N^2$ $c$-number matrix,  which
rationally depends   on the spectral parameters and satisfies the
Yang-Baxter equation (YBE) :
\begin{equation}
 R_{12}(\lambda -\mu )~R_{13}(\lambda )~R_{23}(\mu )~=~
R_{23}(\mu )~R_{13}(\lambda )~R_{12}(\lambda -\mu ) \, ,
\label {(1.2)}
\end{equation}
where  we have used
 the standard notations like $R_{12}(\lambda - \mu ) =
R(\lambda - \mu )  \otimes {\bf 1} $.

However it may be noticed that,
 in spite of their  wide ranged applications,
  Yangian algebras are usually defined through only  one deformation
  parameter dependent rational solutions of YBE (\ref {(1.2)}).
 For example, the rational  $R$-matrix which leads to
  $Y(gl_N)$ Yangian  algebra is  given by [7]
\begin{equation}
R(\lambda ) ~~=~~
 \lambda ~\sum_{i,j=1}^N e_{ii}\otimes e_{jj} ~+~
h ~ \sum_{i,j=1}^N e_{ij}\otimes e_{ji} ~ ,
\label {(1.3)}
\end{equation}
where $(e_{ij})_{kl}= \delta_{ik}\delta_{jl}$ and $h$ is the
single  deformation parameter. At $ h\rightarrow 0 $ limit,
the related  $Y(gl_N)$ algebra reduces to a subalgebra of
infinite dimensional  $gl(N)$ Kac-Moody
algebra containing its non-negative modes.
 On the other hand it is worth noting
that there exists a large class of
 multiparameter dependent   quantum groups [8-10], which
might  be considered as some  deformations of
finite dimensional Lie groups and
can be defined through the spectral parameterless limit of QYBE (\ref
{(1.1)}).
 The $R$-matrix associated with  such  multiparameter dependent
deformation of  $GL(N)$ group is  given  by
\begin{equation}
R~ ~=~~ q~ \sum_{i=1}^N e_{ii} \otimes e_{ii}~ +~ \sum_{i\neq j}~
e^{i\alpha_{ij} }~
e_{ii} \otimes e_{jj}~ +~ (q-q^{-1})~\sum_{i<j }e_{ij}\otimes e_{ji}~,
\label {(1.4)}
\end{equation}
where one takes
 $\alpha_{ij}= - \alpha_{ji} \, $. Apart from the `quantum'
parameter $q$, the above $R$-matrix evidently depends on ${N(N-1)\over
2}$ number of
independent parameters $\alpha_{ij}$. So it should be
interesting to see whether these multiple deformation parameters can
also be incorporated in the rational
$R$-matrix structure (\ref{(1.3)}) and investigate the corresponding
modification of $Y(gl_N)$ Yangian algebra. In sec.2 of this article
we shall  briefly review the  construction of
standard $Y(gl_N)$ algebra and subsequently
demonstrate that it is indeed possible to deform such algebraic
structure through these  ${N(N-1) \over 2}$
number of  extra  parameters.

It may be recalled that  the
 multiparameter dependent  quantum groups
(\,e.g.  $GL_{p,q}(2)$\,) are in general  difficult to
realise  through  generators of corresponding single parameter
dependent  version (\,e.g. $GL_q(2)$\,). However we surprisingly find that,
 in contrast to the
case of usual quantum groups,  there exists a nonlinear
realisation of present multiparameter dependent extension  of
$Y(gl_N)$ algebra  through the generators of
standard $Y(gl_N)$ algebra. By using such  realisation, along with
the known representations for standard $Y(gl_N)$
[11-13], one may  easily
build up  the representations for multideformed
$Y(gl_N)$ algebra.
Moreover, that   realisation interestingly allows  us
to construct a new   $\left( 1+ {N(N-1) \over 2} \right ) $ number of
 deformation parameter dependent coproduct  for
original   $Y(gl_N)$ algebra. We discuss these results in sec.3 of
this paper and make some concluding remarks in  sec.4.
\vskip .5cm
\section{Multiparameter  dependent
extension  of $Y(gl_N)$ Yangian algebra }
\setcounter{equation}{0}

Let us briefly review
 some properties of standard  $Y(gl_N)$  algebra,
before exploring  its multiparameter deformed version.
This  $Y(gl_N)$  algebra can be  generated  by
substituting the
  rational $R(\lambda )$-matrix (\ref {(1.3)})
  to QYBE (\ref {(1.1)}) :
\begin{equation}
(\lambda - \mu )~ \left [~ T^{ij}(\lambda)~,~ T^{kl}(\mu)~
\right ]~~=~~h
{}~\left (~ T^{kj}(\mu )~
T^{il}(\lambda )~ -~T^{kj}(\lambda )~T^{il}(\mu )~ \right ),
\label {(2.1)}
\end{equation}
where $T^{ij}(\lambda )$ are operator valued elements of matrix
$T(\lambda )$.  If we assume the usual
 analyticity property of $T(\lambda )$
 and  asymptotic condition $T(\lambda )
\rightarrow 1$ at $\lambda \rightarrow \infty$ [7], then
 the  elements $T^{ij}(\lambda )$   can
 be expanded in powers of $\lambda $ as
\begin{equation}
T^{ij}(\lambda)~~=~~\delta_{ij}~+~h~ \sum _{n=0}^{\infty}~{t_n^{ij} \over
\lambda^{n+1}}~.
\label {(2.2)}
\end{equation}
Plugging   the above   expansion in eqn. (\ref {(2.1)}) and comparing
from its both sides
the coefficients of same powers in spectral parameters, one
may easily  express
 the $Y(gl_N)$ algebra through the modes $t_n^{ij}$ :
$$
  \left [~ t_0^{ij} \, ,~ t_n^{kl} \, \right ] ~=~\delta_{il}~ t_n^{kj}
{}~- ~ \delta_{kj} ~t_n^{il}~,~~~
 \left[ ~t_{n+1}^{ij} \, , \,  t_m^{kl}~ \right ]
{}~-~ \left [~ t_{n}^{ij}~ ,~ t_{m+1}^{kl}~ \right ] ~=~
h ~ ( ~t_m^{kj}\, t_n^{il} ~- ~ t_n^{kj}\, t_m^{il} ~)~. \eqno (2.3a,b)
$$
Moreover,
with  the help of induction procedure, it  is not difficult to verify that
the $Y(gl_N)$ algebra (2.3a,b) can  be equivalently written  through
a  single relation  given by
\begin{eqnarray}
{}~~\left [~t_n^{ij}~,~ t_m^{kl}~\right ] ~~=~~\delta_{il}~t_{n+m}^{kj}
{}~-~ \delta_{kj}~t_{n+m}^{il} ~+~
 h ~\sum_{p=0}^{n-1}~\left ( ~ t_{m+p}^{kj}~ t_{n-1-p}^{il}
- ~ t_{n-1-p}^{kj}~t_{m+p}^{il}~ \right )~.
\nonumber
\hskip .6 cm (2.3c)
\end{eqnarray}
\addtocounter{equation}{1}
Notice that at the limit $h \rightarrow 0$,
eqn. (2.3c) reduces to a
subalgebra of $gl(N)$ Kac-Moody algebra containing its  non-negative
modes. Consequently, this Yangian algebra might be considered as some nonlinear
 deformation
of Kac-Moody algebra through the parameter $h$. The Casimir operators
for $Y(gl_N)$ algebra  may also be   obtained  by  constructing
the  corresponding quantum determinant ($\delta $) [1,7]. Moreover,
by imposing the restriction  $\delta =1 $, one can  generate the
nonabelian $Y(sl_N) $ algebra form this $Y(gl_N) $ algebra.
It may be observed further that
Yangian algebras form a class of Hopf algebra,  for which the operations like
 coproduct $(\Delta )$, antipode $(\epsilon )$ can  be defined [1].
For example, by using the expression $~\Delta T(\lambda ) ~=~ T(\lambda )
{\buildrel \otimes \over .} T(\lambda )$ as well as  eqn.
(\ref {(2.2)}), it is easy to write down
the coproducts for  all modes of $T(\lambda )$ :
\begin{equation}
\Delta t_0^{ij} ~=~ {\bf 1} \otimes t_0^{ij} ~+~ t_0^{ij} \otimes
{\bf 1}~,~~~
\Delta t_n^{ij} ~=~ {\bf 1} \otimes t_n^{ij} ~+~ t_n^{ij} \otimes
{\bf 1}~+~ h\, \sum_{p+q = n-1}\,  \sum_{k=1}^N ~ t_p^{ik} \otimes t_q^{kj}~,
\label {(2.4)}
\end{equation}
where $n \in [1, \infty ] $ and $p,~q \in [0,\infty ] ~$.

The  mode expansion (2.2) can  be equivalently expressed through another
set of generators $J_n^a$ as
\begin{equation}
T^{ij}(\lambda )
{}~=~ \delta_{ij} ~+~ \sum_{n=0}^\infty ~ { 1\over \lambda^{n+1} }
  ~ \big \{  ~
J_n^0 ~\delta_{ij} ~+~ \sum_{a=1}^{N^2-1} J_n^a ~(t^a)_{ij} ~
\big  \} ~, \label {(2.5)}
\end{equation}
where $t^a$s are traceless Hermitian matrices
 corresponding to the fundamental
representation of $su(N)$ Lie algebra:
$[~t^a , t^b~] ~=~f^{abc}\, t^c$.
Again,  the $Y(gl_N)$ algebra might  be written
 in terms of new  modes $J_n^0$ and $J_n^a$,  by
substituting (\ref {(2.5)}) to  eqn.(\ref {(2.1)})
and comparing the coefficients of spectral parameters. In particular,
the modes  $ J_0^a $ and $ J_1^b $ satisfy
the commutation  relations
\begin{equation}
{}~~[~J_0^a~ ,~ J_0^b~]~=~f^{abc} J_0^c ~,~~~~~[~J_0^a~,~J_1^b~]~=~f^{abc}
{}~J_1^{c} ~.~
\label {(2.6)}
\end{equation}
Furthermore, it turns out that,
all higher level generators of $Y(gl_N)$ or $Y(sl_N)$ algebra
can be realised consistently  through these 0 and 1-level generators,
provided they satisfy   a few   additional Serre relations [1,2].
 The representation theory of these Yangian
algebras  has also been studied in the literature [11-13]
and found to be closely connected with  the degenaracies of wave functions
corresponding to some quantum integrable spin chains [4,7].

Now for extending the above  discussed
 $Y(gl_N)$ algebra to the multiparameter
deformed case, it is necessary  to   generalise first the
form of  related rational  $R(\lambda )$-matrix (\ref {(1.3)}).
For this purpose  we  may
notice that the spectral parameterless
 $ R$-matrix (\ref {(1.4)}), associated with
multiparameter dependent   deformation of $GL(N)$ group,
 satisfies the Hecke like condition [14]
\begin{equation}
 R ~-~ {\tilde  R} ~~=~~ (~q-q^{-1}~)~{\cal P} \, , \label {(2.7)}
\end{equation}
where $~ {\tilde R }~
=~{\cal P}  R^{-1}{\cal  P}$  and $~ {\cal P}~=~\sum\limits_{i,j=1}^N
e_{ij} \otimes e_{ji}~$ being the permutation operator.
So by  following Jones' Yang-Baxterisation prescription
related to Hecke algebra, one can
easily construct a  spectral parameter dependent
 $ R(\lambda)$ matrix  as [15]
$$ R (\lambda)~~=~~q^{\lambda \over h}~ R ~- ~
q^{-{\lambda \over h}}~ {\tilde  R }~,$$
which would be a solution of YBE (1.2).
Substituting
the explicit form of $ R$-matrix (\ref {(1.4)}) to  the above expression
one subsequently   gets
\begin{eqnarray}
 R(\lambda) ~~=~~
 \left ( ~ q^{(1+{\lambda \over h} )} -
  q^{ - (1+{\lambda \over h} ) }~ \right )~ \sum _{i=1}^N
 e_{ii}\otimes e_{ii}
 ~+~ (~q^{{\lambda \over h}} - q^{-{\lambda \over h}}~) ~ \sum _{i\neq j}
  ~e^{ i \alpha_{ij}} ~  e_{ii}\otimes e_{jj}  \nonumber  \\
+ ~ (q-q^{-1})~ \left \{ ~ q^{{\lambda \over h}} ~
\sum _{i<j}~ e_{ij}\otimes e_{ji} ~+~ q^{-{\lambda \over h}} ~\sum _{i>j}
{}~ e_{ij}\otimes e_{ji}~ \right \}~. \label {(2.8)}
\end{eqnarray}
It may be noticed that
this type of  deformation parameter dependent,
trigonometric solution of YBE
was  considered earlier  by Perk and Schultz in the context of solvable
vertex models [16]. However,  the present way of deriving such solution
reveals    its  close connection  to
 multideformed quantum groups.
If  we multiply the $ R(\lambda)$ matrix (\ref {(2.8)})
 by a factor $h / ( q-q^{-1}) $ and then  take the
$q \rightarrow 1$ limit,  that would finally  yield a
 rational solution of YBE  given by
\begin{equation}
R(\lambda)~~=~~\lambda ~\left ( ~ \sum_{i=1}^N
 ~e_{ii}\otimes e_{ii}~+~
\sum _{i\neq j}~ e^{i\alpha_{ij}}~
 e_{ii}\otimes e_{jj}~\right ) ~ +~ h~{\cal P}~.  \label {(2.9)}
\end{equation}
Apart from  parameter $h$, the above
  $R(\lambda )$-matrix evidently
depends on ${N(N-1) \over 2}$  number of
antisymmetric deformation parameters
$\alpha_{ij}$. Moreover, at the limit $\alpha_{ij}=0$ for
all $i,j$, it reduces
to the $R(\lambda )$-matrix (\ref {(1.3)}) related to standard $Y(gl_N)$
Yangian  algebra.

Having found the multiparameter dependent rational
$R(\lambda )$-matrix (\ref {(2.9)}), we
like to explore at present
 the corresponding modification of $Y(gl_N)$ algebra.
So we substitute $R(\lambda )$-matrix  (\ref {(2.9)})
 to  QYBE (\ref {(1.1)}) and  find that
the  previous eqn.(\ref{(2.1)})  gets deformed to
\begin{equation}
(\lambda - \mu)~
\left \{ ~\phi_{ik}~ T^{ij}(\lambda) \,  T^{kl}(\mu)~-~
\phi_{jl}~ T^{kl}(\mu) \,  T^{ij}(\lambda)~\right \}~=~h~\left (
 ~T^{kj}(\mu) \, T^{il}(\lambda)~-
 ~T^{kj}(\lambda) \, T^{il}(\mu)~\right ) \, ,
\label {(2.10)}
\end{equation}
 where the notation $\phi_{ij} = e^{i\alpha_{ij}} $ has been used.
 Next we attempt  to express
the above multideformed algebra through the modes of $T^{ij}(\lambda )$.
 However in this context one curiously observes
  that the usual mode expansion (\ref {(2.2)}),
related to asymptotic condition $T(\lambda ) \rightarrow 1$
at $  \lambda  \rightarrow \infty$, is no longer consistent with the
  algebra (\ref {(2.10)}). To avoid this problem
  we propose a modification of
mode expansion (\ref {(2.2)})  as
\begin{equation}
 T^{ij}(\lambda) ~~=~~ \delta _{ij} \,\tau^{ii}~+~ \sum_{n=0}^{\infty }~
 {{\hat t}^{\,ij}_n \over \lambda^{n+1}} ~, \label {(2.11)}
\end{equation}
where $\tau^{ii}$s  ( $i \in [1,N]$ )
 are some additional nontrivial generators.
By substituting this modified expansion to  eqn. (\ref {(2.10)}) and
equating from its both sides
   the coefficients of
 same  powers in  $\lambda ,~\mu, $
it is not difficult to obtain  the following independent  relations
 among the modes $\tau^{ii}, ~{\hat t}^{\, ij}_n$ :
\begin{eqnarray*}
     \left[\, \tau^{ii} \, , \, \tau^{jj} \, \right ] ~=~0~,~~~~
\tau^{ii} \,{\hat t}^{ \, jk}_n ~=~
{\phi _{ik} \over \phi_{ij}}~ {\hat t}^{\,jk}_n \, \tau^{ii}~,~~
\hskip 3.2 cm   &&(2.12a,b) ~
\\
     \phi_{ik}~{\hat t}^{\, ij}_0 \, {\hat t}^{ \, kl}_n
{}~-~\phi_{jl}~ {\hat t}^{ \, kl}_n  \, {\hat t}^{ \, ij}_0
{}~~=~~
\delta_{il}~{\hat t}^{ \, kj}_n \, \tau^{ii}~-~\delta_{kj}
{}~\tau^{kk} \, {\hat t}^{ \, il}_n~,~~ ~ \hskip 2 cm &&~~~(2.12c)
 \\
 \left [ ~\phi_{ik}~{\hat t}^{ \, ij}_{n+1} \, {\hat t}^{ \, kl}_m
{}~ - ~\phi_{jl}~
{\hat t}^{ \, kl}_m \, {\hat t}^{ \, ij}_{n+1}~ \right ] ~-~
\left [ ~\phi_{ik}~{\hat t}^{ \, ij}_n \,
{\hat t}^{ \, kl}_{m+1}~-~\phi_{jl}~
{\hat t}^{ \, kl}_{m+1}{\hat t}^{ \, ij}_n~ \right ] \hskip 1.7 cm
&& \\
 \hskip 8.9 cm ~=~   h~\left ( ~{\hat t}^{ \, kj}_m \,
{\hat t}^{ \, il}_n  ~-~  {\hat t}^{ \, kj}_n \, {\hat t}^{ \, il}_m~
\right ) ~.&&~~~(2.12d)
\end{eqnarray*}
With the help of  induction procedure, subsequently we find  that
the equations
(2.12c)  and (2.12d) can also be expressed through
a single relation given by
\begin{eqnarray*}
{}~\phi_{ik}~{\hat t}^{ \, ij}_n \, {\hat t}^{ \, kl}_m ~ - ~
\phi_{jl}
 ~{\hat t}^{ \, kl}_m \, {\hat t}^{ \, ij}_n~~=~~
\delta_{il}~{\hat t}^{ \, kj}_{n+m} \,  \tau^{ii}~-~\delta_{kj}
{}~ \tau^{kk} \, {\hat t}^{ \, il}_{n+m}~  \hskip 4.75 cm   \\
 \hskip 7 cm + ~h~
\sum_{p=0}^{n-1}~ \left ( ~{\hat t}^{ \, kj}_{m+p}\,
 {\hat t}^{ \, il}_{n-1-p}~-~{\hat t}^{ \, kj}_{n-1-p}
 \, {\hat t}^{ \, il}_{m+p}~ \right ) ~.~~~~ (2.12e)
\end{eqnarray*}
\addtocounter{equation}{1}
Comparing with  standard
 $Y(gl_N)$ algebra (2.3),
we notice  that the   relations   (2.12a-e)  (\,$Y_{m}(gl_N)$ algebra\,)
 depend on some extra generators $\tau^{ii}$
as well as deformation parameters $\phi_{ij}~( = e^{ i \alpha_{ij}} ) $ .
Moreover
from    eqn. (2.12b) it is evident  that these  $\tau^{ii}$s
do not commute with all other  elements  of the  deformed  algebra.
This fact
  clearly indicates  why the
previous  mode expansion (\ref{(2.2)}), recoverable from
(\ref{(2.11)}) by fixing   $\tau^{ii}=1$,
leads to inconsistencies in  the present  context.
 However  in the particular limit
  $\phi_{ij} = 1 $ for all $i,j$, one can
consistently put  $\tau^{ii}=1$ in the relations (2.12a-e). Consequently,
   at this  limit our  multiparameter dependent $ Y_{m}(gl_N ) $  algebra
reduces  to  its single parameter dependent version (2.3).

In analogy with   the case
of standard Yangian algebra,
  one  can define the coproduct  for
  $Y_{m}(gl_N)$  algebra (2.12)  by using
 the relation $\Delta T(\lambda )~=~
T(\lambda ){\buildrel  \otimes \over .} T(\lambda )$
and the modified  mode expansion (\ref {(2.11)}) :
\begin{eqnarray}
\Delta \tau^{ii} ~=~ \tau^{ii} \otimes \tau^{ii}~,~~~~~
\Delta {\hat t}_0^{\, ij} ~=~ \tau^{ii} \otimes
 {\hat t}_0^{\, ij} ~+~ {\hat t}_0^{\, ij} \otimes \tau^{jj}~,
\hskip .8 cm  \nonumber \\
\Delta {\hat t}_n^{ \, ij} ~~=~~ \tau^{ii} \otimes
 {\hat t}_n^{ \, ij} ~+~ {\hat t}_n^{ \, ij} \otimes \tau^{jj}~
+~ h~  \sum_{p+q=n-1} ~\!  \sum_{k=1}^N  ~
 {\hat t}_p^{ \, ik} \otimes  {\hat t}_q^{\, kj}~,
  \label {(2.13)}
\end{eqnarray}
where $n \in [1, \infty ] $ and $p,~q \in [0,\infty ] ~$.
Notice that the above relations
  do not explicitly  depend on  extra
parameters $\alpha_{ij}$ and reduce to the  coproduct of
$Y(gl_N)$ (2.4)
under the substitution $\tau^{ii}=1$.
Now it is interesting to enquire  whether the present
  $Y_{m}(gl_N)$  algebra  can also be realised through some 0 and 1-level
generators satisfying the Serre relations. Moreover, it might
be of physical relevence to build up  the representations for  this
multiparameter deformed Yangian algebra. However,  rather than directly
approaching to these problems,
 we shall construct in the  following
 a realisation of
  $Y_{m}(gl_N)$  algebra through the generators of
$Y(gl_N)$ algebra.  The existence of such realisation
   would automatically imply that the
  $Y_{m}(gl_N)$  algebra  can also be realised
through  0 and 1-level  generators of original
 $Y(gl_N)$  algebra.
 Moreover, by using that  realisation
and  known representations of $Y(gl_N)$ algebra [11-13],
one may easily  build up the representations for
  $Y_{m}(gl_N)$  algebra.

\section{ Realisation of
  $Y_{m}(gl_N)$  algebra }
\setcounter{equation}{0}
For constructing a  realisation of
$Y_{m}(gl_N)$  algebra (2.12) through the generators of
$Y(gl_N)$ satisfying  (2.3),  let us
introduce first another associative  algebra
 (\,${\tilde Y}(gl_N) $\,) which contains  the modes
$\tau^i$ and $t_{n}^{ij}$ :
\begin {eqnarray*}
{}~~[~ \tau^i \, , \, \tau^j~] ~=~ 0~, ~~~ \tau^i t_n^{jk} ~=
{}~ e^{ {i\over 2} ( \alpha_{ik} - \alpha_{ij} )} ~
 t_n^{jk}  \tau^i ~=~ \sqrt { \phi_{ik}\over \phi_{ij} } ~
 t_n^{jk}  \tau^i  ~,~~~~~~~~~(3.1a,b) \\
{}~~~~\left [~t_n^{ij}~,~ t_m^{kl}~\right ] ~=~\delta_{il}~t_{n+m}^{kj}
- \delta_{kj}~t_{n+m}^{il} ~+~
 h ~\sum_{p=0}^{n-1}~\left (  ~ t_{m+p}^{kj}\, t_{n-1-p}^{il}
{}~ - ~  t_{n-1-p}^{kj}\, t_{m+p}^{il}~ \right ) ~.
{}~~~~~~(3.1c)
\end{eqnarray*}
\addtocounter{equation}{1}
Notice  that the
 above defined ${\tilde Y}(gl_N)$
is in some sense intermediate between
  $Y(gl_N)$ and $Y_{m}(gl_N)$ algebra. On the one hand this
 ${\tilde Y}(gl_N)$ contains the modes $t_n^{ij}$, which
 evidently generate the $ Y(gl_N)$ as a subalgebra. While, on the other
hand,  it has  $N$ number of  extra generators $\tau^i$ which are
much   similar to  the generators $\tau^{ii}$ of
multideformed $Y_m(gl_N)$ algebra. By taking advantage of
 this curious link  provided by ${\tilde Y}(gl_N)$,
 we  shall construct in the following
a realisation of
$Y_{m}(gl_N)$  algebra  through the generators of
${\tilde Y}(gl_N) $. Subsequently  we shall demonstrate that this
intermediate ${\tilde Y}(gl_N) $ algebra, in turn,
  can be realised through the
standard $Y(gl_N)$ generators. Finally, by combining these two realisations,
we would be able to express the $Y_m(gl_N)$ generators
in terms of $Y(gl_N)$ generators.

To begin with  we  observe that the
$Y_m(gl_N)$ algebra (2.12) allows a simple realisation
  through the generators of ${\tilde Y}(gl_N)$ (3.1) as
\begin{equation}
\tau^{ii}~=~ (\tau^i)^2~,~~~~~ {\hat t}_n^{\, ij}~=~
\exp \, \left( \, -{i\alpha_{ij}\over 2} \, \right)~
\tau^i \tau^j~ t^{ij}_n~. \label {(3.2)}
\end{equation}
For checking
 the validity of this  statement, one may consider the case
of eqn. (2.12c) as an example. By substituting in it the realisation
(\ref{(3.2)}) and then using the
commutation relations  (3.1a,b), it can be brought in the
form
\begin{equation}
{\cal F}~[~ t_0^{ij} ~,~ t_n^{kl}~ ] ~=~{\cal F}~
\left (~\delta_{il}~ t_n^{kj}
{}~- ~ \delta_{kj} ~t_n^{il}~ \right )~, \label {(3.3)}
\end{equation}
where ${\cal F}~=~
\exp~ [ \, {i\over 2} \, ( \, \alpha_{ik}+\alpha_{jk}+\alpha_{lk}
+\alpha_{jl}+\alpha_{li}+\alpha_{ji} \, ) \, ]~\tau^i \tau^j \tau^k \tau^l$.
Now with the help of
 eqn. (3.1c) one can easily verify  that the relation
(\ref{(3.3)}) is indeed satisfied by the generators of ${\tilde Y}(gl_N)$.
 It is worth noting  that the emergence
of common factor ${\cal F}$ from both sides of eqn. (\ref {(3.3)})
plays a  crucial  role in the above outlined proof.
In a similar way we can check  the validity of
realisation (\ref{(3.2)}) for all
algebraic  relations appearing in  (2.12).

Next, we attempt   to construct a
 realisation of ${\tilde Y}(gl_N)$ algebra through
 the generators $t_n^{ij}$ associated with
    $Y(gl_N)$ algebra (2.3). Since the
 ${\tilde Y}(gl_N)$ algebra  already contains $Y(gl_N)$ as a subalgebra,
all we need in this case   is
 to express the extra  generators $\tau^i$ as some  functions of
 modes $t_n^{ij}$ such that the relations (3.1a,b) would be satisfied.
So  we make an ansatz for these  $\tau^i$s   in the form
\begin{equation}
\tau^i~~=~~ \exp~\left (
 ~ i\sum_{l=1}^N~ \beta_{il} \, t_0^{ll}~ \right )~,
\label {(3.4)}
\end{equation}
 where $\beta_{il}$ are some yet undetermined constants.
Due to   relations $ ~[~t_0^{ii},~t_0^{jj}~]~=~0$, which are some
particular  cases of eqn. (2.3a), it is evident
that the above defined
$\tau^i$s are commuting among themselves for arbitrary $\beta_{il}$.
Furthermore,  one may
  observe from eqn. (2.3a) that
the modes $t_n^{ij}$ behave like ladder operators
with respect to $t_0^{ii}$ :
\begin{equation}
[~t_0^{ii}~,~t_n^{ij}~]~=~- ~t_n^{ij}~,~~~
[~t_0^{ii}~,~t_n^{ji}~]~=~t_n^{ji}~,~~~
[~t_0^{ii}~,~t_n^{jk}~]~=~0~,
\label {(3.5)}
\end{equation}
 where $ i\neq j \neq k ~$.
By using the relation (\ref {(3.5)}) it is not difficult to verify that
the ansatz (\ref {(3.4)})  would satisfy
eqn. (3.1b) provided one  takes  $~\beta_{il}~=~{ \alpha_{il} \over 2}~ $.
Consequently,  for this particular value of $\beta_{il} \,$ the expression
(\ref {(3.4)}) yields  a realisation of ${\tilde Y}(gl_N)$ algebra
through the generators of $Y(gl_N)$ algebra.
Finally,  by combining the relations  (\ref{(3.2)}) and
 (\ref{(3.4)}),  we  obtain a  realisation of
 multideformed $Y_m(gl_N)$ algebra (2.12) through the generators
 of single parameter dependent $Y(gl_N)$ algebra (2.3) :
\begin{equation}
\tau^{ii} ~=~ \exp~\left(~
 i \sum_{l=1}^N \alpha_{il}\,  t_0^{ll}~\right)  ~,~~~~~
{\hat t}_n^{\, ij} ~=~ \exp ~\left ( ~ -
 { i\alpha_{ij}   \over 2 } ~+~{i\over 2}~
\sum_{l=1}^N ~ \left ( \alpha_{il} + \alpha_{jl} \right )~t_0^{ll}~ \right )
{}~t_n^{ij}~.
\label{(3.6)}
\end{equation}

With the help of  above
 realisation  and known representations [11-13]
of  $Y(gl_N)$ algebra,  one   can easily
construct  the representations for $Y_m(gl_N)$ algebra. Furthermore,
by using such realisation,  it is also possible to express all higher
level generators of $Y_m(gl_N)$   through the  0 and 1-level
 generators of $Y(sl_N)$ algebra   satisfying the usual Serre
relations. However,  one may consider
  eqn. (\ref {(3.6)}) as a nonlinear
transformation  which removes the deformation parameters
$\alpha_{ij}$ from the   $Y_m(gl_N)$ algebra and maps it to
 single parameter dependent   $ Y(gl_N)$ algebra.
But   it is   interesting to observe  further  that,
with the help of same  transformation (\ref{(3.6)}),
 the coproduct of $Y_m(gl_N)$ (\ref{(2.13)}) induces
a novel  $\left ( \, 1+ {N(N-1) \over 2 } \, \right) $ number of deformation
 parameter dependent  coproduct for
standard  $Y(gl_N)$ algebra :
\begin{equation}
\Delta t_0^{ij} ~ = ~  K_{ij} \otimes t_0^{ij} ~ + ~  t_0^{ij} \otimes
 K_{ji} ~,~~~
\Delta t_n^{ij} ~ =~   K_{ij} \otimes t_n^{ij} ~ + ~  t_n^{ij} \otimes
 K_{ji} ~ +  ~
h\, \sum_{p+q = n-1} \,
 \sum_{k=1}^N ~\Gamma_{ijk}~ t_p^{ik} \otimes t_q^{kj} \, ,
\label {(3.7)}
\end{equation}
where we have used the notation
$$~K_{ij}~=~\exp \, \left  (  ~{i\over 2} \, \sum\limits_{l=1}^N \,
( \, \alpha_{il} - \alpha_{jl} \, ) ~ t_0^{ll} ~ \right  )~,~~~~
 {\Gamma}_{ijk}~=~ e^{ \,
 {i\over 2}\, \left (\, \alpha_{ij} + \alpha_{jk} + \alpha_{ki} \,
\right ) }
{}~K_{kj} \otimes K_{ki} ~.
$$
 By applying  eqn. (3.5) it is easy to
see that the operators
$K_{ij}$ satisfy the relations like
\begin{equation}
K_{ik} \,  K_{kj} ~=~ K_{ij}~, ~~~~
t_0^{kl}\, K_{ij} ~=~ e^{ \, {i\over 2} \, ( \, \alpha_{ik } +
\alpha_{kj} + \alpha_{jl} + \alpha_{li} \, ) \, }~
  K_{ij}\,  t_0^{kl}~.
\label{(3.8)}
\end{equation}
With the help of above relations one can also directly verify that the
multiparameter dependent coproduct (\ref {(3.7)}) is
  consistent with
$Y(gl_N)$ algebra (2.3). Moreover, at the limit
 $\alpha_{ij} = 0$
for all $i,\, j,$  eqn.  (\ref {(3.7)}) reproduces  the known
coproduct (\ref {(2.4)}) of $Y(gl_N)$.
 Thus it turns out   that,  though
the  transformation (\ref{(3.6)}) might be  used to remove
the  deformation parameters $\alpha_{ij}$
 from  $Y_m(gl_N)$ algebra,
these  deformation parameters interestingly  reappear
in the coproduct of
 resulting $Y(gl_N)$ algebra.

\section{Conclusion}

Here we present a  multiparameter dependent extension of
 $Y(gl_N)$ Yangian algebra,
 in analogy with the case of  multideformed quantum groups.
To achieve this, we first Yang-Baxterise
 a braid group representation associated with
   multideformed version of $GL_{q}(N)$ quantum group and
    take  the necessary
 $q\rightarrow 1$ limit,  for   obtaining  a rational $R$-matrix
which depends on $\left (\, 1+ {N(N-1) \over 2} \, \right ) $
number of deformation parameters. Quantum Yang-Baxter
equation corresponding to such  rational $R$-matrix
interestingly   yields a multiparameter dependent extension of
standard $Y(gl_N)$ algebra (\,$Y_m(gl_N)$ algebra\,).

Next we try   to express  this  $Y_m(gl_N)$ algebra
in  a spectral parameter independent form and find
  that the  usual asymptotic  condition on monodromy matrix
 $T(\lambda )$,  i.e.
 $T(\lambda) \rightarrow 1$ at $ \lambda  \rightarrow \infty ~, $
is no longer appropriate for this purpose.
To overcome this problem,
 we modify such  asymptotic  condition by introducing
$N$ number of extra  generators  $\tau^{ii}$. Consequently,
these extra generators also appear in an intriguing  fashion
  in the
resulting   spectral parameter independent
form of our  $Y_m(gl_N)$ algebra.
Subsequently  we find that there exists
 a  nonlinear realisation of $Y_m(gl_N)$ algebra through the
generators of standard $Y(gl_N)$. By using such realisation
and  known representations of  $Y(gl_N)$ algebra, one can
easily build up  the representations  of deformed $Y_m(gl_N)$ algebra.
Furthermore, this realisation allows us to construct   a
novel  $\left ( 1 + { N(N-1) \over 2 } \right )$ number of deformation
 parameter dependent coproduct for
standard $Y(gl_N)$ algebra.

As has been remarked earlier, the multiparameter dependent
  Yangian algebra considered here
is actually related to the
 rational  limit of trigonometric $R$-matrix originally proposed by
 Perk and Schultz.  However, a
$q$-analogue of standard Yangian algebra can also be  constructed
from quantum Yang-Baxter equation by directly using a single
 parameter dependent
 trigonometric $R$-matrix [17].
So it should be interesting to explore
 the multiparameter
dependent extension  of such   $q$-analogue of Yangian algebra. Moreover,
as it is well known,
   there exists a class of quantum integrable
 models  whose Hamiltonians and
 other conserved quantities
 can be derived  from    some realisations of standard  Yangian algebra.
So it might be of  physical interest to search for new integrable
models which would be  similarly related
 to the   multiparameter dependent  extension  of Yangain algebra.

\end{document}